\documentclass[aps,prd,twocolumn,a4paper,superscriptaddress]{revtex4-1}
\usepackage{graphicx}
\usepackage{multirow}
\usepackage{latexsym}
\usepackage{hyperref}
\usepackage[british]{babel}
\begin{document}

\newcommand{\be}{\begin{equation}}
\newcommand{\ee}{\end{equation}}
\newcommand{\bq}{\begin{eqnarray}}
\newcommand{\eq}{\end{eqnarray}}
\newcommand{\bsq}{\begin{subequations}}
\newcommand{\esq}{\end{subequations}}
\newcommand{\bc}{\begin{center}}
\newcommand{\ec}{\end{center}}

\title{Extending the velocity-dependent one-scale model for domain walls}

\author{C. J. A. P. Martins}
\email{Carlos.Martins@astro.up.pt}
\affiliation{Centro de Astrof\'{\i}sica da Universidade do Porto, Rua das Estrelas, 4150-762 Porto, Portugal}
\affiliation{Instituto de Astrof\'{\i}sica e Ci\^encias do Espa\c co, CAUP, Rua das Estrelas, 4150-762 Porto, Portugal}
\author{I. Yu. Rybak}
\email[]{Ivan.Rybak@astro.up.pt}
\affiliation{Centro de Astrof\'{\i}sica da Universidade do Porto, Rua das Estrelas, 4150-762 Porto, Portugal}
\affiliation{Instituto de Astrof\'{\i}sica e Ci\^encias do Espa\c co, CAUP, Rua das Estrelas, 4150-762 Porto, Portugal}
\affiliation{Faculdade de Ci\^encias, Universidade do Porto, Rua do Campo Alegre 687, 4169-007 Porto, Portugal}
\author{A. Avgoustidis}
\email[]{Anastasios.Avgoustidis@nottingham.ac.uk}
\affiliation{School of Physics and Astronomy, University of Nottingham, University Park, Nottingham NG7 2RD, United Kingdom}
\author{E. P. S. Shellard}
\email[]{E.P.S.Shellard@damtp.cam.ac.uk}
\affiliation{Centre for Theoretical Cosmology, Department of Applied Mathematics and Theoretical Physics, Wilberforce Road, Cambridge CB3 0WA, United Kingdom}

\date{2 February 2016}

\begin{abstract}

We report on an extensive study of the evolution of domain wall networks in Friedmann-Lema\^{\i}tre-Robertson-Walker universes by means of the largest currently available field-theory simulations. These simulations were done in $4096^3$ boxes and for a range of different fixed expansion rates, as well as for the transition between the radiation and matter eras. A detailed comparison with the velocity-dependent one-scale (VOS) model shows that this cannot accurately reproduce the results of the entire range of simulated regimes if one assumes that the phenomenological energy loss and momentum parameters are constants. We therefore discuss how a more accurate modeling of these parameters can be done, specifically by introducing an additional mechanism of energy loss (scalar radiation, which is particularly relevant for regimes with relatively little damping) and a modified momentum parameter which is a function of velocity (in analogy to what was previously done for cosmic strings). We finally show that this extended model, appropriately calibrated, provides an accurate fit to our simulations.

\end{abstract}
\maketitle

%%%%%%%%%%%%%%%%%%%%%%%%%%%%%%%%%%%%%%%%%%%%%%%%%%%%%%%%%%%%%%%%%%%%%%%%%%%%%%%%%%
\section{Introduction}

It is generally accepted that phase transitions occurred during the early stages of the evolution of the Universe. Among their possible consequences is the production of topological defects through the Kibble mechanism~\cite{Kibble}. Two-dimensional topological defects (domain walls) are tightly constrained, unless they are very light or they decay soon after formation, since otherwise they would dominate the energy density of the universe, in disagreement with observations~\cite{Zeldovich}. On the other hand, one-dimensional objects (cosmic strings) are in principle more benign, although they are also subject to increasingly strong constraints~\cite{Planck2013}. Nonetheless, cosmic strings could play an important role as a relic of fundamental theories of the early Universe, such as brane inflation scenarios~\cite{SarangiTye,PolchinskiCopelandMyers} or supersymmetric grand unified theories (GUT)~\cite{JeannerotRocherSakellariadou}.

To understand the observational effects of the presence of topological defects, a quantitative understanding of the evolution of their networks is essential. Such quantitative analytic models were first obtained for cosmic strings \cite{MartinsShellard,MartinsShellard2}, and subsequently for domain walls~\cite{AvelinoMartinsOliveira}. Meanwhile, the latter can be more easily simulated numerically at higher spatial resolution and dynamic range. For this reason, in addition to their intrinsic relevance, domain walls also provide a useful testbed for the evolution of cosmic strings and superstrings \cite{Testbed}. Still, the velocity-dependent one-scale model (VOS) for domain walls is currently less developed than its cosmic string counterpart. Here we take advantage of recent improvements in hardware and computing power to improve this situation.

Specifically, we build upon the work done in~\cite{LeiteMartins,LeiteMartinsShellard} and carry out an extensive set of high-resolution field theory simulations of domain wall networks using the PRS algorithm~\cite{PRS}. Compared to this earlier work our simulations are both larger ($4096^3$ boxes, the largest currently available) and span a more diverse set of conditions, including simulations with fixed expansion rates (radiation era, matter era and 10 other expansion rates) as well as, for the first time for domain walls, series of simulations that accurately span the radiation-matter transition. This extended high-resolution dataset enables us to further calibrate and significantly improve the analytic model, as was previously done for strings \cite{VOS1,VOS2}.

%%%%%%%%%%%%%%%%%%%%%%%%%%%%%%%%%%%%%%%%%%%%%%%%%%%%%%%%%%%%%%%%%%%%%%%%%%%%%%%%%%

\section{The Standard VOS Model for Domain Walls}

The analytic VOS model for domain wall network evolution was first obtained from arguments on energy conservation in~\cite{AvelinoMartinsOliveira}. Later it was shown that the same result can be reached from a microscopic description~\cite{AvelinoSousa}. We will revisit and clarify this microscopic approach, and further extend it to shed light on the \textit{momentum parameter} $k$ for the wall network (to be rigorously defined below).

\begin{figure}[!]
\begin{center}
\includegraphics[width=3in]{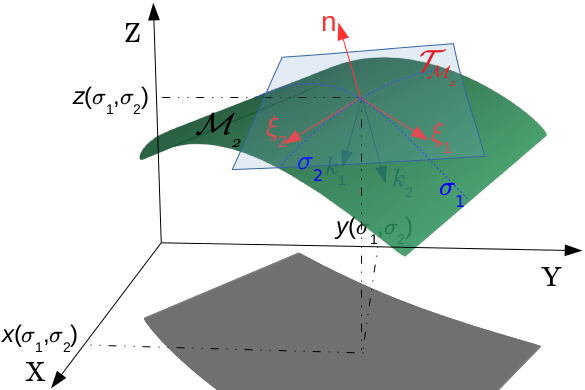}
\end{center}
\caption{\label{fig:TanVect} The wall surface $\mathcal{M}_2$ parametrized by two parameters, $\sigma_1$ and $\sigma_2$.}
\end{figure}

The wall surface $\mathcal{M}_2$ can be parametrized by two parameters, $\sigma_1$ and $\sigma_2$. As a result, the wall evolution is described by the vector $x^{\mu}(\sigma_1,\sigma_2, \tau)$, where we identified $\sigma_0=\tau$.\footnote{Throughout this work Greek indices $\mu, \nu$, $\lambda$ run from $0$ to $3$ denoting space-time coordinates, Latin indices $i,k,l$ run from $1$ to $3$ denoting spatial coordinates, and $a,b,c$ run from $0$ to $2$ denoting coordinates on the domain wall worldvolume.} If the function $x^{\mu}(\sigma_1,\sigma_2, \tau)$ is smooth, it is possible to parametrize the wall surface in such a way that two tangential vectors will be orthogonal 
\begin{equation}
   \label{Orthogs}
   \partial_{\sigma_1}x^{\mu} \partial_{\sigma_2}x_{\mu} \equiv x^{\mu}_{,1} x_{\mu,2} = 0\,.  
\end{equation}
Moreover, we can require that the velocity of the wall $\partial_{\tau} x^{\mu} \equiv \dot{x}^{\mu}$ can be only normal to the tangent 
surface $\mathcal{T_{M}}_2$ (cf. Fig. \ref{fig:TanVect}).

%%%%%%%%%%%%%%%%%%%%%%%%%%%%%%%%%%%%%%%%%%%%%%%%%%%%%%%%%%%%%%%%%%%%%%%%%%%%%%%%%%
%%%%%%%%%%%%%%%%%%%%%%%%%%%%%%%%%%%%%%%%%%%%%%%%%%%%%%%%%%%%%%%%%%%%%%%%%%%%%%%%%%

To derive the wall equation of motion we start from the worldvolume (Dirac) action, which has the form
\begin{equation}
   \label{Lagr}
   S = -\int \mathcal{L} d^3 \sigma = - \sigma_{w} \int \sqrt{\gamma} d^3 \sigma,  
\end{equation}
where $\sigma_{w}$ is a constant mass per unit area, $\gamma_{ab} = g_{\mu \nu} x^{\mu}_{,a} x^{\nu}_{,b}$ is the induced metric, 
$\gamma = \frac{1}{3!} \epsilon^{ab} \epsilon^{cd} \gamma_{ac} \gamma_{bd}$ is its determinant, 
$x^{\mu}_{,a} = \frac{\partial x^{\mu}}{\partial \sigma^{a}}$, $\epsilon^{ab} $ is the Levi-Civita symbol, and $\mathcal{L}$ is the Lagrangian density.

To obtain equations of motion for a domain wall from Eq. (\ref{Lagr}), it is useful to use the following equality
\begin{equation}
   \label{Equality}
   d \mathcal{L} = \frac{1}{2} \sqrt{\gamma} \gamma^{ab} d \gamma_{ab}\,, 
\end{equation}
from which one can obtain
\begin{widetext}
\begin{equation}
   \label{EqOfMot}
\frac{\partial \mathcal{L}}{\partial x^{\lambda}} -  \partial_{c} \left( \frac{\partial \mathcal{L}}{\partial x^{\lambda}_{,c}} \right) = 0 = \frac{1}{2} \sqrt{\gamma} \gamma^{ab} g_{\mu \nu, \lambda} x^{\mu}_{,a} x^{\nu}_{,b} -  \partial_{c} \left( \sqrt{\gamma} \gamma^{ab} g_{\mu \lambda} x^{\mu}_{,a} \delta^{c}_{b} \right)\,,
\end{equation}
\begin{equation}
   \label{Stress-Energ}
T^{\mu \nu} \sqrt{-g} \equiv -2 \frac{\delta S}{\delta g_{\mu \nu}} = \sigma_{w} \int \sqrt{\gamma} \gamma^{ab} x^{\mu}_{,a} x^{\nu}_{,b} \delta^4(x^{\rho} - x^{\rho}(\sigma^{c}) ) d^2 \sigma d \tau\,,
\end{equation}
where $g$ is the determinant of the metric $g_{\mu \nu}$. The energy of the wall in that case is 
\begin{equation}
   \label{Energ}
E = \sigma_{w} a(\tau) \int \sqrt{\gamma} \gamma^{00} d^2 \sigma = \sigma_{w} a^2(\tau) \int \varepsilon d^2 \sigma\,.  
\end{equation}

Let us now define the metric $g_{\mu \nu}$ as the FLRW metric with conformal time $a(\tau) d\tau = dt$
\begin{equation}
   \label{FLRW}
    ds^2 = a^2(\tau) \left( d\tau^2 - dl^2 \right),
\end{equation}
where $a(\tau)$ is the scale factor and $dl^2 = dx^2 + dy^2 + dz^2 $.
Then the equation of motion~(Eq. \ref{EqOfMot}) can be rewritten as 
\begin{equation}
   \label{EqOfMot2}
 \frac{\dot{a}}{a} \delta_{0 \lambda} \sqrt{\gamma} \gamma^{ab} \gamma_{ab} -  \partial_{c} \left( \sqrt{\gamma} \gamma^{ab} g_{\mu \lambda} x^{\mu}_{,a} \delta^{c}_{b} \right) =0\,.
\end{equation}
\end{widetext}

Let us redefine the coordinates $\sigma_1$ and $\sigma_2$ to $s_1$ and $s_2$ in such way that $|\frac{\partial x^i}{\partial s_{\alpha}}|^2 = 1$ ($\alpha=1,2$). This means that derivatives will be changed in the following way
\begin{equation}
   \label{Paramchange}
    \frac{\partial x^i}{\partial \sigma_{\alpha}} = |x^i_{,\alpha}| \frac{\partial x^i}{\partial s_{\alpha}},
\end{equation}
(no summation over $\alpha$). In these new coordinates, it is possible to introduce an orthonormal basis (refer to Fig.~\ref{fig:TanVect}): $\xi^i_{\alpha}=\frac{\partial x^{i}}{\partial s_{\alpha}}$, and $\mathrm{n}^i=\frac{\dot{x}^{i}}{|\dot{x}^{i}|}$. Consequently, the zeroth component of Eq.~(\ref{EqOfMot2}) ($\lambda=0$) can be written as
\begin{equation}
   \label{EqOfMotEnerg}
   \dot{\varepsilon} + 3 \frac{\dot{a}}{a} \varepsilon \dot{x}^i \dot{x}_i = 0\,.
\end{equation}
The spatial part $\left( \lambda = i \right)$ of Eq.~(\ref{EqOfMot2}) contracted with the vector $\mathrm{n}_i$ has the form
\begin{equation}
   \label{EqOfMotSpatial}
\ddot{x}^{i} \mathrm{n}_{i} + 3 \frac{\dot{a}}{a} \dot{x}^{i} \mathrm{n}_{i} \left( 1 - \dot{x}^{i} \dot{x}_i \right) = \left( 1 - \dot{x}^i \dot{x}_i \right) k^{i}_{1} \mathrm{n}_{i} + \left( 1 - \dot{x}^i \dot{x}_i \right) k^{i}_{2} \mathrm{n}_{i}\,,
\end{equation}
where $k^i_{\alpha} = \frac{\partial\xi^i_{\alpha}}{\partial s_{\alpha}} $.

The scalar products $k^{i}_{\alpha} \mathrm{n}_{i}$ project the curvatures corresponding to $\sigma_1$ and $\sigma_2$ along the normal vector $\mathrm{n}^i$. It should be noted that $k^i_{\alpha} = \frac{a}{R_{\alpha}} u^i_{\alpha}$, where $u^i_{\alpha}$ are unit vectors and $R_{\alpha}$ are the radii of curvature for $\sigma_1$ and $\sigma_2$, respectively. 

Now it is possible to obtain averaged equations, using the same strategy that was used in \cite{MartinsShellard}. One introduces two macroscopic (averaged) quantities, the energy density
\begin{equation}
   \label{EnergyDens}
    \frac{E}{V}=\rho=\frac{\sigma_{w} a^2}{V} \int \varepsilon d^2 \sigma
\end{equation}
and the root-mean-squared (rms) velocity
\begin{equation}
   \label{RMSvel}
    \upsilon^2 = \frac{\int \dot{x}^2 \varepsilon d^2 \sigma}{\int \varepsilon d^2 \sigma}\,,
\end{equation}
and can thus average Eqs. (\ref{EqOfMotEnerg}-\ref{EqOfMotSpatial}), obtaining

\begin{eqnarray}
   \label{AveragedEqSyst}
   \frac{d \rho}{d t} & = & - H \rho  \left( 1 + 3 \upsilon^2 \right)\,, \nonumber \\
   \frac{d \upsilon}{d t} & = & \left( 1 - \upsilon^2 \right) \left( \frac{K_1 + K_2}{L} - 3 Hv \right)\,,
\end{eqnarray}
where $t$ is a physical time, and $H= \frac{1}{a} \frac{d a}{d t} $ is the Hubble parameter, and we made the assumption that curvature radii have the same averaged value and are equal to the correlation length: $R_1=R_2=L$. The $K_1$ and $K_2$ parameters are curvature/momentum parameters. The component $K_1$ can be written as
\begin{equation}
   \label{defK1}
   K_1 = u_1^i n_i\,,
\end{equation}
suitably averaged over the network, with an analogous definition for $K_2$. As a first approximation they may be assumed to be constants, but later on in this work we will address how they may depend on the velocity $\upsilon$. 

%%%%%%%%%%%%%%%%%%%%%%%%%%%%%%%%%%%%%%%%%%%%%%%%%%%%%%%%%%%%%%%%%%%%%%%%%%%%%%%%%%

\section{Simulations and preliminary calibration} \label{Simulation}

An evolving wall network loses energy because of possible intersections and the creation of sphere-like objects that eventually collapse. This energy loss mechanism can be added to Eqs. (\ref{AveragedEqSyst}) by analogy to what was originally done by Kibble for cosmic strings \cite{Kibble2}. This has the form
\begin{equation}
   \label{Chopping}
    \frac{d \rho_{loss}}{d t} =- c_w \upsilon \frac{\rho}{L}\,,
\end{equation}
where $c_w$ is a constant which we will call (by analogy to the cosmic strings case) the \textit{chopping} parameter. 

Taking into account this energy loss term, we can rewrite Eqs. (\ref{AveragedEqSyst}) in terms of the correlation length $L=\frac{\sigma_{w}}{\rho}$, as follows
\begin{eqnarray}
   \label{AveragedEqSyst2}
    \frac{dL}{dt} & = & (1 + 3 \upsilon^2)H L + c_w \upsilon \,, \nonumber \\
    \frac{d\upsilon}{dt}  & = & (1-\upsilon^2) \left( \frac{k_w}{L} - 3 H \upsilon \right)\,,
\end{eqnarray}
where we further defined $k_w = K_1 + K_2$ as the \textit{momentum} parameter, which we will initially consider as a constant.

For a FLRW universe expanding as a power law, $a \propto t^{\lambda}$, Eqs. (\ref{AveragedEqSyst2}) have the asymptotic scaling solution $L=\epsilon t$ and $\upsilon = v_0$, where $\lambda$, $\epsilon$ and $v_0$ are constants~\cite{AvelinoMartinsOliveira}. The two phenomenological parameters $c_w$ and $k_w$ can then be expressed as
\begin{equation}
   \label{NetParameters1}
    k_w = 3 \lambda \epsilon v_0 \,,
\end{equation}
\begin{equation}
    c_w v_0 = \epsilon \left[ 1 - \lambda \left( 1 + 3 v_0^2 \right) \right]\,.
   \label{NetParameters2}
\end{equation}

It should be noted that the right hand-sides of Eqs.~(\ref{NetParameters1}-\ref{NetParameters2}) are general expressions for the momentum parameter and energy loss mechanisms that can be measured directly and independently from simulations. For this purpose, we should obtain the asymptotic values of the quantities $\epsilon$ and $v_0$ from our simulations.

\begin{figure}[!]
\begin{center}
\includegraphics[width=3.5in]{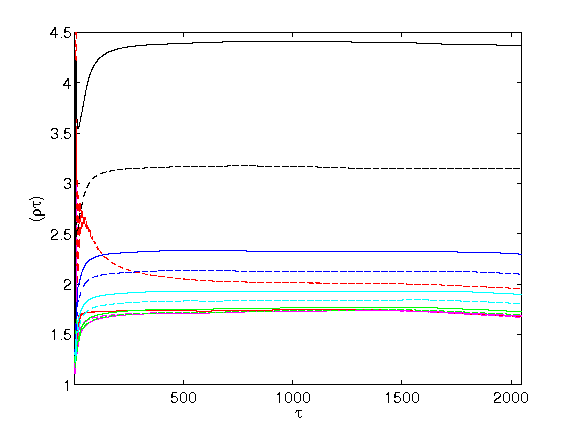}
\includegraphics[width=3.5in]{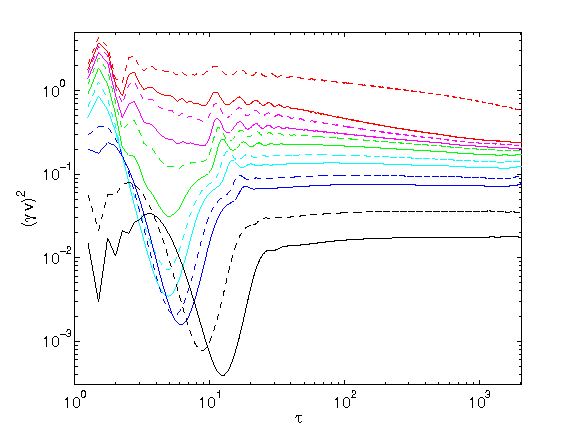}
\end{center}
\caption{\label{lambdaboxes}The evolution of the dimensionless density ($\rho\tau$, top panel) and the rms speed $(\gamma v)^2$ (where $\gamma$ is the Lorentz factor, bottom panel) in $4096^3$ domain wall simulations with different expansion rates, from $\lambda=1/10$ (red dashed, corresponding to the highest velocity and lowest density) to $\lambda=19/20$ (black solid, corresponding to the highest density and lowest velocity).}
\end{figure}

Building upon the work done in Refs.~\cite{LeiteMartins,LeiteMartinsShellard}, we have carried out field theory simulations of the simplest (single-field) domain wall networks in a FLRW background. The equations of motion for the scalar field $\phi$, adopting the Press, Ryden and Spergel procedure~\cite{PRS}, can be written in terms of conformal time $\tau$ as
\begin{equation}
   \label{Field-theoryEq}
   \frac{\partial^2 \phi}{\partial \tau^2} + 3 \frac{\mathrm{d} \ln a}{\mathrm{d} \ln \tau} \frac{\partial \phi}{\partial \tau} - \frac{\partial^2 \phi}{\partial x^i \partial x_i} = - \frac{\partial V}{\partial \phi}\,.
\end{equation}
Relevant numerical parameters are $\phi_0 = \pm1$ for the minima of the potential, while the maximum of the potential is $V_0 = \pi^2 / 2 W_0^2$ (where $W_0=10$ is the initial wall thickness in grid units). All these are similar to the ones used in earlier simulations~\cite{PRS,LeiteMartins,LeiteMartinsShellard}. Relative to earlier works our simulations have three key advantages
\begin{itemize}
\item We used a faster and more memory-efficient version of our earlier WALLS code \cite{LeiteMartins,LeiteMartinsShellard}, optimized for the Intel Xeon Phi architecture.
\item This optimization allows us to increase the box size (and therefore the spatial resolution and dynamic range). Specifically, we ran several series of $4096^3$ simulations on the COSMOS supercomputer, thus gaining a factor of 8 in volume and a factor of 2 in dynamic range as compared to Ref. \cite{LeiteMartinsShellard}. Each simulation starts with $\tau_i=1$ and is stopped when the horizon becomes half the box size ($\tau_f=2048$), ensuring that the periodic boundary conditions of the simulation boxes do not affect the results. Each such simulation requires 1 Tb of memory and takes about 3.7 hours of wall clock time to run on 512 CPUs.
\item We explore a much larger range of fixed expansion rates, including radiation era ($\lambda=1/2$), matter era ($\lambda=2/3$) and 10 other expansion rates, ranging from $\lambda=1/10$ to $\lambda=19/20$. (Additionally we also simulated universes during the transition from radiation to matter, to which we will return below.) For each choice of expansion rate we have carried out 10 simulations with different (random) initial conditions: although each of the 10 choices was made randomly, the same 10 choices were used for each of the simulated expansion rates. (This ensures that any differences can be solely ascribed to the different expansion rates.) Unless otherwise stated, the results presented in what follows correspond to the average of each set of 10 runs. Figure \ref{lambdaboxes} illustrates the results of these constant expansion rate simulations.
\end{itemize}

Note that our choice of initial conditions will lead to large energy gradients in the early timesteps of the simulation, and the network needs some time (which is proportional to the wall thickness) to wash away these initial conditions. This implies that in many grid points the field will go over the top of the potential to get into the other minimum, transiently leading to a relatively small average velocity (the more so the faster the expansion rate), which is clearly visible in the early timesteps in the bottom panel of Fig. \ref{lambdaboxes}---note that $\tau=10$ is the light-crossing time for walls of the average thickness being simulated. This erasing of initial conditions is done in a quasi-coherent way at the various points in the box, leading to the damped oscillations in the average velocity that are also visible  in the bottom panel of Fig. \ref{lambdaboxes} (though in this case they are clearer for the slower expansion rates, corresponding to weaker damping).

%%%%%%%%%%%%%%%%%%%%%%%%%%%%%%%%%%%%%%%%%%%%%%%%%%%%%%%%%%%%%%%%%%%%%%%%%%%%%%%%%%
%%%%%%%%%%%%%%%%%%%%%%%%%%%%%%%%%%%%%%%%%%%%%%%%%%%%%%%%%%%%%%%%%%%%%%%%%%%%%%%%%%

In practice, since the simulations are evolved in conformal time, the quantity we measure is the conformal correlation length divided by conformal time, which can be straightforwardly related to the physical time quantities
\begin{equation}
   \label{ConformalXi}
   \frac{\xi_c}{\tau}=(1-\lambda)\frac{L}{t}=(1-\lambda)\epsilon\,.
\end{equation}
Similarly for the velocity we measure $\gamma v$ (or, more precisely, $(\gamma v)^2$), where $\gamma = \frac{1}{\sqrt{1-v^2}}$ is the Lorentz factor \cite{LeiteMartins}. In order to identify accurate asymptotic values, we should find the simulation dynamic range when $\xi_c/\tau$ and $\gamma v$ have already reached the asymptotic behavior and the simulation box still has enough walls for robust statistics (towards the end of each simulation only a few long walls remain, resulting in comparatively poor statistics). After some tests, we conservatively defined the region $\tau = 500 - 1500$ in which our simulations are generally well behaved for all expansion rates (specifically, they are in scaling solutions without significant fluctuations).

Once this region is specified, the averaged values of $\xi_c/\tau$ and $\gamma v$ can be obtained. These results are presented in Table~\ref{TableMeasure}. Together with values of $\xi_c/\tau$ and $\gamma v$ we also list the scaling exponents $\nu$ and $\mu$, quantifying convergence to the attractor scaling solution. These are defined as
\begin{equation}
   \label{Scalingmu}
   \frac{1}{\xi_c} \propto \tau^{\mu}
\end{equation}
\begin{equation}
   \label{Scalingnu}
   \gamma v \propto \tau^{\nu}\,,
\end{equation}
so for a scaling network these exponents should be numerically consistent with $\mu=-1$ and $\nu=0$. As expected, one finds that the convergence to the scaling solution is faster for faster expansion rates (corresponding to a larger damping term in the wall equations of motion). Indeed, the $\nu$ diagnostic shows that for the slowest expansion rate we have simulated ($\lambda=1/10$) the network has not converged to the scaling behavior and, as a result, it cannot be used for further analysis. In fact this is also qualitatively clear from a simple visual inspection of Fig. \ref{lambdaboxes}.

%%%%%%%%%%%%%%%%%%%%%%%%%%%%%%%%%%%%%%%%%%%%%%%%%%%%%%%%%%%%%%%%%%%%%%%%%%%%%%%%%%
\begin{table*}[ht]
\centering
\caption{Scaling properties of numerical simulations for domain wall networks with different expansion rate $\lambda$ in the range $\tau = \left( 500 - 1500 \right)$. See the main text for the definition of the various parameters.}
\label{TableMeasure}
\begin{tabular}{c | c c | c c | c c}
\hline
$\lambda$ & $\mu$ & $\nu$ & $\xi_c/\tau$ & $\gamma v$ & $k_w$ & $c_w$  \\ [0.5ex]
\hline
1/10 & $-1.020 \pm 0.005$ & $-0.147 \pm 0.001$ & $0.496 \pm 0.016$  & $0.867 \pm 0.040$ & $0.108 \pm 0.004$ & $0.65 \pm 0.03 $ \\
1/5 & $-0.992 \pm 0.005$ & $-0.085 \pm 0.003$ & $0.575 \pm 0.020$ & $0.514 \pm 0.017$ & $0.20 \pm 0.01$ & $1.06 \pm 0.04$ \\
1/4 & $-0.984 \pm 0.005$ & $-0.066 \pm 0.003$ & $0.578 \pm 0.020$ & $0.489 \pm 0.015$ & $0.25 \pm 0.01$ & $1.06 \pm 0.04$  \\
1/3 & $-0.984 \pm 0.005$ & $-0.057 \pm 0.003$ & $0.580 \pm 0.021$ & $0.467 \pm 0.013$ & $0.37 \pm 0.02$ & $1.00 \pm 0.04$ \\
2/5 & $-0.983 \pm 0.005$ & $-0.054 \pm 0.004$ & $0.577 \pm 0.021$ & $0.449 \pm 0.014$ & $0.47 \pm 0.03$ & $0.94 \pm 0.04 $\\
1/2 & $-0.989 \pm 0.005$ & $-0.046 \pm 0.004$ & $0.568 \pm 0.019$ & $0.418 \pm 0.012$ & $0.66 \pm 0.04$ & $0.81 \pm 0.04$  \\
3/5 & $-0.996 \pm 0.004$ & $-0.039 \pm 0.004$ & $0.545 \pm 0.018$ & $0.379 \pm 0.012$ & $0.87 \pm 0.05$ & $0.67 \pm 0.05$\\
2/3 & $-1.000 \pm 0.004$ & $-0.032 \pm 0.004$ & $0.519 \pm 0.015$ & $0.348 \pm 0.011$ & $1.02 \pm 0.05$ & $0.56 \pm 0.06$\\
3/4 & $-1.003 \pm 0.003$ & $-0.026 \pm 0.004$ & $0.470 \pm 0.012$ & $0.302 \pm 0.008$ & $1.22 \pm 0.06$ & $0.41 \pm 0.07$ \\
4/5 & $-1.006 \pm 0.003$ & $-0.021 \pm 0.003$ & $0.430 \pm 0.009$ & $0.269 \pm 0.007$ & $1.34 \pm 0.05$ & $0.31 \pm 0.06$ \\
9/10 & $-1.006 \pm 0.002$ & $\, 0.003 \pm 0.003$ & $0.316 \pm 0.004$ & $0.190 \pm 0.004$ & $1.59 \pm 0.05$ & $0.11 \pm 0.06$  \\
19/20 & $-0.997 \pm 0.001$ & $\, 0.008 \pm 0.002$ & $0.227 \pm 0.002$ & $0.133 \pm 0.002$ & $1.70 \pm 0.03$ & $0.03 \pm 0.04$ \\
\hline
\end{tabular}
\end{table*}
%%%%%%%%%%%%%%%%%%%%%%%%%%%%%%%%%%%%%%%%%%%%%%%%%%%%%%%%%%%%%%%%%%%%%%%%%%%%%%%%%%

Using the asymptotic values, one can obtain $\epsilon = \frac{\xi_{c}}{\tau (1-\lambda)} $ and the velocity $v_0$ for each expansion rate. By inserting $\epsilon$ and $v_0$ into Eqs.~(\ref{NetParameters1}-\ref{NetParameters2}) one numerically obtains the momentum and chopping parameters. The values thus obtained for each expansion rate are also listed in Table~\ref{TableMeasure}. It is noteworthy that, with the exception of the $\lambda=1/10$ case, $k_w$ increases monotonically with $\lambda$, while $c_w$ correspondingly decreases.

%%%%%%%%%%%%%%%%%%%%%%%%%%%%%%%%%%%%%%%%%%%%%%%%%%%%%%%%%%%%%%%%%%%%%%%%%%%%%%%%%%
\begin{figure}[!]
\begin{center}
\includegraphics[width=3.5in]{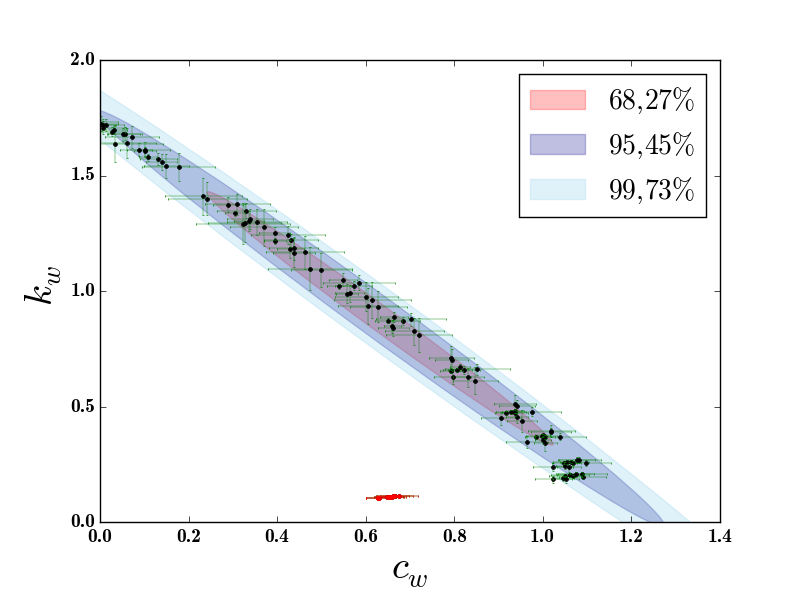}
\includegraphics[width=3.5in]{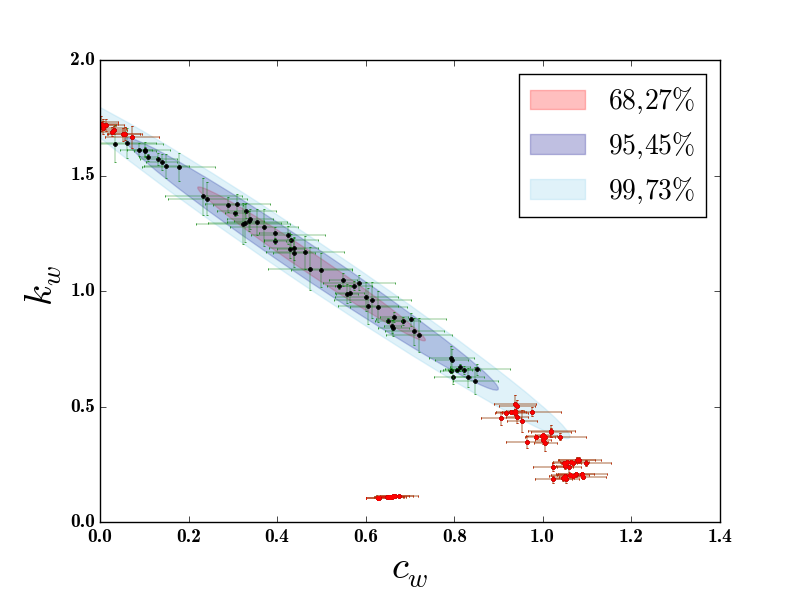}
\end{center}
\caption{\label{fig:CwandKw} The likelihood contours for the VOS model with constant parameters $c_w$ and $k_w$, for all scaling expansion rates $0.2\le\lambda\le0.95$ (top panel) and for the restricted range $0.5\le\lambda\le0.9$ (bottom panel). Each point with error bars in the plot presents asymptotic values from one simulation. The black dots denote the simulations used in the fit, and the red dots the simulations not used. The slowest expansion rate data was not used in either case: it has a manifestly different behavior because the simulations did not reach the asymptotic scaling behavior.}
\end{figure}
%%%%%%%%%%%%%%%%%%%%%%%%%%%%%%%%%%%%%%%%%%%%%%%%%%%%%%%%%%%%%%%%%%%%%%%%%%%%%%%%%%

For comparison with previous work \cite{LeiteMartins,LeiteMartinsShellard}, it is interesting to carry out a joint analysis of the data (except the $\lambda=1/10$ case), and determine the best-fit values for these phenomenological parameters if one imposes that they should have the same constant value for all epochs. The results of this analysis are shown in the top panel of Fig.~\ref{fig:CwandKw}, and the following best-fit parameters and uncertainties were found $c_w = 0.63 \pm 0.36$, $k_w = 0.88 \pm 0.51$.

It should be noted that in this analysis only the statistical errors were taken into account, because it's not possible to determine the systematic error for these simulations. Such errors would include effects such as the PRS approximation \cite{PRS} and the choice of the range of conformal times in which to do the fits and numerically measure the scaling parameters. However, it is expected that the systematic error becomes smaller for higher expansion rates (since the code is then more robust \cite{LeiteMartins}), just as the statistical error does. Therefore, while the full errors can be larger than presented here, they are broadly expected to have the correct behavior as a function of the expansion rate: higher expansion rates provide more accurate data. As a result, this set of data can be reliably used to calibrate and extend the VOS model.

If we compare the likelihood contours in the top panel of Fig.~\ref{fig:CwandKw} with the same plot from Ref.~\cite{LeiteMartinsShellard} (which only had data from three expansion rates, $\lambda=1/2,2/3,4/5$), it is seen that  they are statistically consistent, but in our case the error bars are significantly larger (as is the reduced chi-squared for the fit). As a comparison, if we repeat the analysis using only the simulations in the range $0.5\le\lambda\le0.9$ we find $c_w = 0.48 \pm 0.24$, $k_w =1.12 \pm 0.31$; the results of this analysis are shown in the bottom panel of Fig.~\ref{fig:CwandKw}. The error bars become smaller and the agreement with Ref.~\cite{LeiteMartinsShellard} is even better. This implies that assuming $c_w$ and $k_w$ to be constants is not accurate enough for these simulations, if one aims to model a broad range of expansion rates.

\begin{figure}[!]
\begin{center}
\includegraphics[width=3.5in]{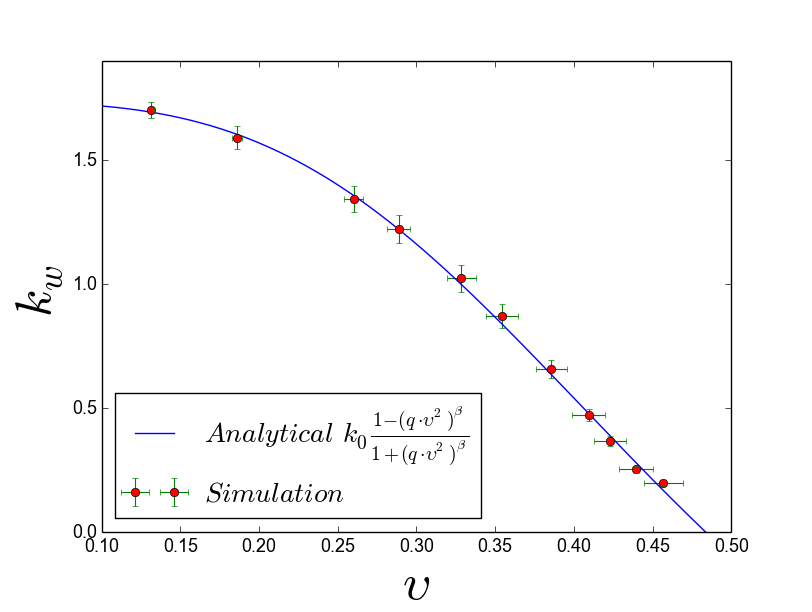}
\includegraphics[width=3.5in]{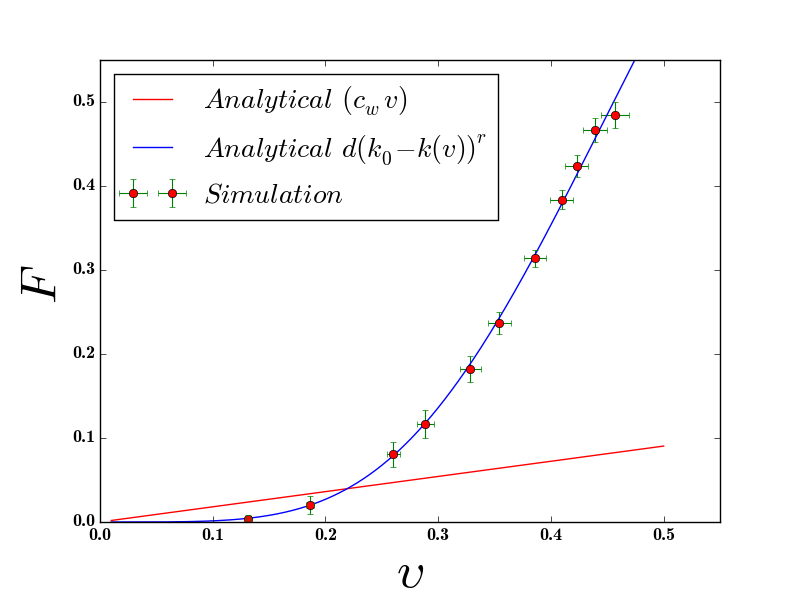}
\end{center}
\caption{\label{fig:KwandF}Momentum parameter $k(v)$ (top panel) and energy loss function $F(v)$ (bottom panel), as numerically determined from the right-hand side of Eqs.~(\ref{NetParameters1}-\ref{NetParameters2}). The red line in the energy loss plot is a linear function of the rms velocity $c_w v$ fitted for high $\lambda$ (hence low velocity). The blue lines are from the extended analytic model, using phenomenological forms of the momentum parameter~(Eq. \ref{MomentumParameter}) and energy loss due to scalar radiation~(Eq. \ref{ScalarRadiation+Chopping}) with the following best-fit parameters $d=0.28$, $r=1.30$, $\beta = 1.69$, $k_0 = 1.73$ and $q=4.27$, discussed in the text.}
\end{figure}

More explicitly this can be shown by plotting the right-hand sides of Eqs.~(\ref{NetParameters1}-\ref{NetParameters2}) in terms of the velocity---cf. Fig.~\ref{fig:KwandF}. For the first of these (top panel), the right-hand side describes the behavior of the momentum parameter $k(v)$ for different expansion rates, while for the second one (bottom panel) it describes an energy loss function which we will denote $F(v)$. For constant $k_w$ and $c_w$, Eqs.~(\ref{NetParameters1}-\ref{NetParameters2}) would imply a constant value for the first plot and a linear function for the second one. Data from our simulations show that this is not the case. As a final, more straightforward check, Table~\ref{TableMeasure} also lists the numerically inferred $c_w$ and $k_w$ for each expansion rate. Hence, the momentum parameter should depend on velocity, and the chopping parameter is not sufficient for describing the energy losses.

%%%%%%%%%%%%%%%%%%%%%%%%%%%%%%%%%%%%%%%%%%%%%%%%%%%%%%%%%%%%%%%%%%%%%%%%%%%%%%%%%%
%%%%%%%%%%%%%%%%%%%%%%%%%%%%%%%%%%%%%%%%%%%%%%%%%%%%%%%%%%%%%%%%%%%%%%%%%%%%%%%%%%

%%%%%%%%%%%%%%%%%%%%%%%%%%%%%%%%%%%%%%%%%%%%%%%%%%%%%%%%%%%%%%%%%%%%%%%%%%%%%%%%%%

\section{Extending the VOS model}

We now describe how to extend the analytic VOS model, by more accurately modeling the momentum parameter and the energy loss term. 

\subsection{Momentum parameter}

The momentum parameter can be estimated in an analogous way to what was done for cosmic strings in Ref.~\cite{MartinsShellard2}. As we saw, the momentum parameter in our VOS wall model is given by $k_w=K_1+K_2$ (refer to Sec. II). The component $K_1$ was previously defined in Eq. (\ref{defK1}), suitably averaged over the network, and a similar definition applies to $K_2$.  We now need to estimate this scalar product in terms of the velocity $\upsilon$.  As can be seen in Fig.~\ref{fig:TanVect}, there is an orthonormal basis $\left\lbrace \xi_1^i, \xi_2^i, n^i  \right\rbrace$, and therefore we can decompose the vector
\begin{equation}
   \label{decomp}
   u_1^i = A n^i + B \xi_2^i\,;
\end{equation}
note that vector $\xi_1^i$ is orthogonal to $u_1^i$. Therefore $u_1^i n_i = A$ and $u^i_1 u_{1i} = A^2 + B^2 = 1$.

When the expansion rate is slow, the velocity squared tends to some maximal value $1/q$ and perturbations on the wall surface increase. Since the wall surface is highly perturbed in that regime, the averaged value of $u_1^i n_i$ goes to zero ($A \rightarrow 0$). In the opposite limit when the rate of expansion is fast, the velocity squared tends to zero, and perturbations on the wall surface are very small. As a result, the scalar product $u_1^i n_i$ goes to some value $k_0/2$. The same considerations apply for $K_2$.
 
Hence, $k(v)$ should reach some value $k_0$ when the velocity is zero and tend to zero when the velocity squared is $1/q$. In that case $k(v)$ can be written similarly to the momentum parameter of the string network~\cite{MartinsShellard2}
\begin{equation}
   \label{MomentumParameter}
k(v)=k_{0} \frac{1-\left( q  \upsilon^2  \right)^{\beta} }{1 + \left(q \upsilon^2 \right)^{\beta} },
\end{equation}
where $\beta$, $k_0$ and $q$ are unknown parameters.

At this point there is one difference between the string and wall cases: there are no non-trivial analytic solutions for walls (like the helicoidal solution for strings) that can be used to infer exact values of $q$, $k_0$ and $\beta$, as it was done for strings. Consequently, it is only possible to impose physical restrictions on these parameters. The constant $k_0$ characterizes the maximum value of the momentum parameter: it is positive, but cannot be bigger than $2$. The parameter $1/q$ is an averaged maximal velocity for the wall network. Similarly to what was done for strings in~\cite{Kibble2}, using the general expression for the $n$-dimensional topological defect dynamics~\cite{AvelinoSousa}, it can be shown that the maximal possible velocity is $v_{max}^2=\frac{n}{n+1}$. For walls this is $v_{w}^2 = 2/3$, as expected, but this result requires a set of assumptions that need not be satisfied. In that case the maximal averaged velocity of the network can be smaller (but not larger). As a result we have
\begin{equation}
   \label{Klimits}
0 < \frac{1}{q} \leq v_{w}^2\,.
\end{equation}
Other than these general physical constraints, these parameters must be calibrated numerically. Fortunately, the resolution of our simulations is high enough to enable this calibration, as we will show below.

\subsection{Energy loss mechanisms}

The modification of the momentum parameter described above is not sufficient to account for the mismatch between the simulation data and the analytic prediction for the energy losses. We should also improve the modeling of the latter for a better description of the wall network evolution. In addition to the chopping mechanism, another significant contribution to energy losses is expected to be from scalar radiation. Moreover, one may expect it to be proportionally more important (compared to the chopping mechanism) for slower expansion rates.

Energy loss due to scalar radiation was considered in Ref.~\cite{VachaspatiEverettVilenkin}. It was shown that the uniformly moving wall does not radiate. Only perturbations on the wall surface produce scalar radiation. We have already estimated the level of perturbations in the momentum parameter expression $k(v)$. The maximal value $k_0$ corresponds to the minimal rms velocity and hence to minimal perturbations on the wall surface. Conversely the case when the momentum parameter is zero corresponds to a maximal rms velocity and a maximally perturbed surface. It looks reasonable to anticipate that the amount of radiation is proportional to the surface perturbations. As a result, we can introduce a modified analytic description of energy losses 
\begin{equation}
   \label{ScalarRadiation+Chopping}
F(v) = c_wv + d[k_0 - k(v)]^{r}\,,
\end{equation}
where $d$ and $r$ are constants. In the maximally perturbed (slow expansion) limit $v^2\rightarrow 1/q$ this behaves as
\begin{equation}
   \label{MaxPert}
F(v) = \frac{c_w}{\sqrt{q}} + dk_0^r\,,
\end{equation}
and we expect the scalar radiation term to be the dominant one. Conversely in the uniform surface (fast expansion) limit we have
\begin{equation}
   \label{MinPert}
F(v) \sim c_w v + d(2k_0)^rq^{\beta r}v^{2\beta r}\,,
\end{equation}
and in this case we expect the chopping term to be more important, and possibly dominate (depending on the values of the free parameters to be calibrated numerically).

Putting together these extensions, the VOS model equations (Eq.\ref{AveragedEqSyst2}) can finally be rewritten as
\begin{eqnarray}
   \label{AveragedEqSyst3}
   \frac{dL}{dt} &=& (1 + 3 \upsilon^2)H L + c_w \upsilon + d[k_0-k(v)]^r ,  \nonumber \\
   \frac{d\upsilon}{dt} &=& (1-\upsilon^2) \left( \frac{k(v)}{L} - 3 H \upsilon \right),
\end{eqnarray}
where $k(v)$ is defined by Eq.~(\ref{MomentumParameter}).

\subsection{Calibrating the extended model}

One can easily confirm that the extended VOS model given by Eqs.~(\ref{AveragedEqSyst3}) possess the same scaling behavior as the original one, given by Eqs.~(\ref{AveragedEqSyst2}). In the extended model we have in principle $6$ undefined parameters that should be determined from numerical simulation data. By using bootstrapping techniques one finds that the chopping parameter is negligibly small in comparison with the contribution from scalar radiation and may be neglected as a first approximation (specifically, we find $c_w=0.00\pm0.01$), while the other five parameters have the following values
\begin{equation}
   \label{fitd}
d=0.28 \pm 0.01
\end{equation}
\begin{equation}
   \label{fitr}
r=1.30 \pm 0.02
\end{equation}
\begin{equation}
   \label{fitbeta}
\beta = 1.69 \pm 0.08
\end{equation}
\begin{equation}
   \label{fitk0}
k_0 = 1.73 \pm 0.01
\end{equation}
\begin{equation}
   \label{fitq}
q=4.27 \pm 0.10\,.
\end{equation}
The scaling solution of Eqs.~(\ref{AveragedEqSyst3}) with the best-fit parameters is also shown in Fig.~\ref{fig:KwandF} by the blue lines. As can be seen, this now provides an excellent agreement with the entire range of numerical simulations.

For comparison, we have also repeated this analysis for the restricted range $0.5\le\lambda\le0.9$, finding
\begin{equation}
   \label{fitd2}
d=0.29 \pm 0.01
\end{equation}
\begin{equation}
   \label{fitr2}
r=1.30 \pm 0.06
\end{equation}
\begin{equation}
   \label{fitbeta2}
\beta = 1.65 \pm 0.12
\end{equation}
\begin{equation}
   \label{fitk02}
k_0 = 1.72 \pm 0.03
\end{equation}
\begin{equation}
   \label{fitq2}
q=4.10 \pm 0.17\,,
\end{equation}
which are fully consistent with the ones obtained for the full range of expansion rates. While this is not entirely surprising (since the fit is dominated by the high expansion rates, for which the statistical uncertainties of the parameters measured in the simulations are smaller) it is supporting evidence for the fact that the model can accurately describe all expansion rates.

We note that in this restricted range the chopping parameter is still negligible ($c_w=0.00\pm0.03$), indicating that much smaller velocities (and therefore faster expansion rates than $\lambda=19/20$) would be needed to make this term comparable to the scalar radiation one. Numerically exploring this ultra-fast expansion regime is an interesting but computationally challenging task which we leave for subsequent work.

%%%%%%%%%%%%%%%%%%%%%%%%%%%%%%%%%%%%%%%%%%%%%%%%%%%%%%%%%%%%%%%%%%%%%%%%%%%%%%%%%%
\begin{figure}[!]
\begin{center}
\includegraphics[width=3.5in]{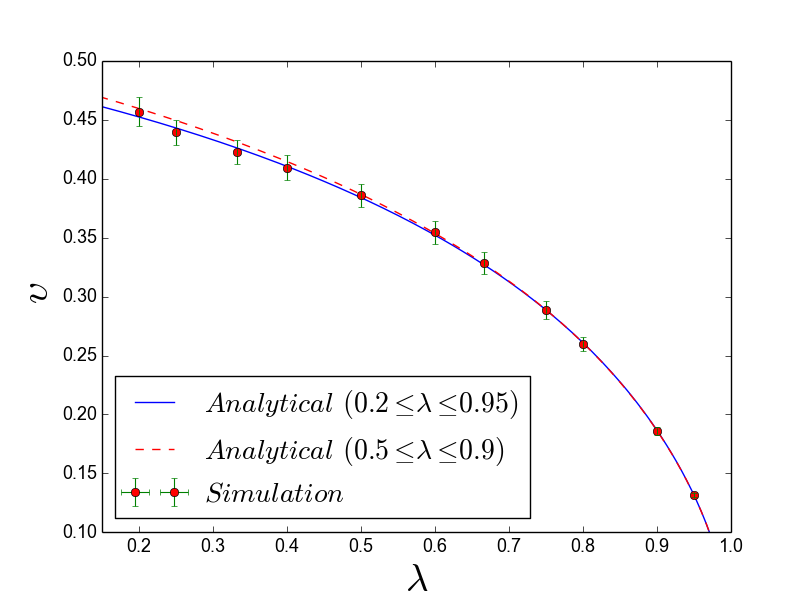}
\includegraphics[width=3.5in]{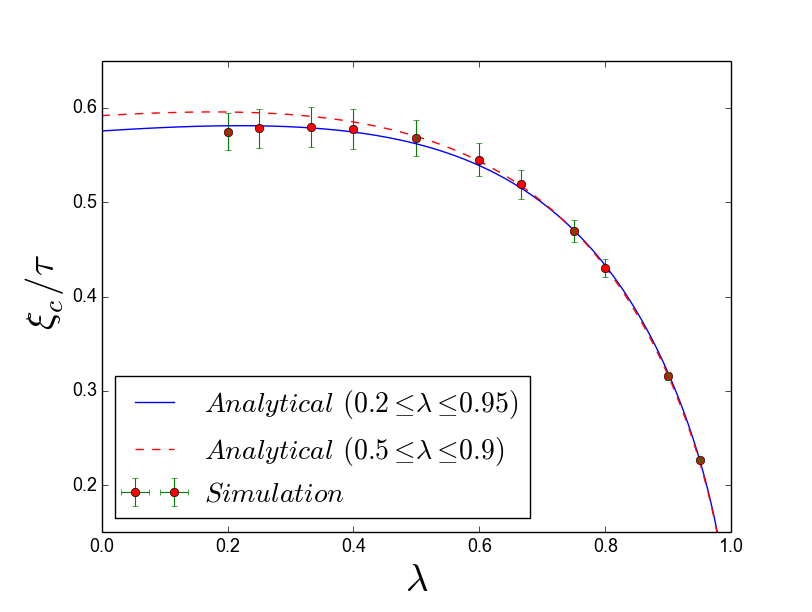}
\end{center}
\caption{\label{fig:v+xi}Velocity $v$ and conformal correlation length divided by conformal time $\xi_c/\tau$ obtained from the model using Eqs. (\ref{AveragedEqSyst3}) with the best-fit parameters described in the text, compared to the data (with statistical error bars) from the numerical simulations for different expansion rates. The solid blue line corresponds to the best-fit parameters for the full range of expansion rates considered while the red dashed one corresponds to the best-fit parameters for the restricted range.}
\end{figure}
%%%%%%%%%%%%%%%%%%%%%%%%%%%%%%%%%%%%%%%%%%%%%%%%%%%%%%%%%%%%%%%%%%%%%%%%%%%%%%%%%%

Alternatively, by solving Eqs. (\ref{AveragedEqSyst3}) in the scaling regime for different expansion rates $\lambda$ with the best-fit parameters determined above we can compare the model predictions with the numerically determined quantities. Figure~\ref{fig:v+xi} displays this comparison (both for the full and restricted ranges of the expansion rate $\lambda$), confirming that the extended model accurately reproduces the simulation data. Moreover, it can be concluded that the main energy loss mechanism for the wall network is generically scalar radiation, with the chopping term only becoming important for fast expansion rates.

%%%%%%%%%%%%%%%%%%%%%%%%%%%%%%%%%%%%%%%%%%%%%%%%%%%%%%%%%%%%%%%%%%%%%%%%%%%%%%%%%%

\section{The radiation-matter transition}

Thus far we tested the model against numerical simulations with a fixed expansion rate $\lambda$. As an additional test, we have carried out analogous field theory simulations of the radiation-matter transition. In this case the scale factor has the following exact analytic expression
\begin{equation}
   \label{ScaleFact}
   \frac{a(\tau)}{a_{eq}}= \left( \frac{\tau}{\tau_{*}} \right)^2 + 2 \left( \frac{\tau}{\tau_{*}} \right)\,,
\end{equation}
where $\tau_{*}=\tau_{eq}/(\sqrt{2}-1)$ and the parameters $a_{eq}$ and $\tau_{eq}$ are constants denoting the scale factor and conformal time at the epoch of equal radiation and matter densities. For illustration purposes we can also calculate an 'effective' expansion rate during the transition
\begin{equation}
 \lambda_{eff}=\frac{2+2\frac{\tau}{\tau_{*}}}{4+3\frac{\tau}{\tau_{*}}}\,;
\end{equation}
as expected this interpolates between the radiation and matter era values.

%%%%%%%%%%%%%%%%%%%%%%%%%%%%%%%%%%%%%%%%%%%%%%%%%%%%%%%%%%%%%%%%%%%%%%%%%%%%%%%%%%
\begin{figure}[!]
\begin{center}
\includegraphics[width=3.5in]{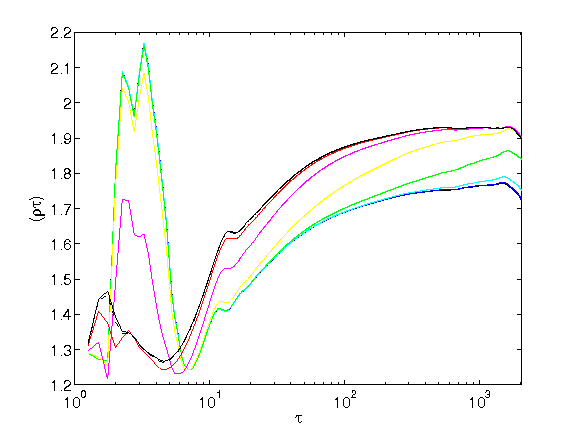}
\includegraphics[width=3.5in]{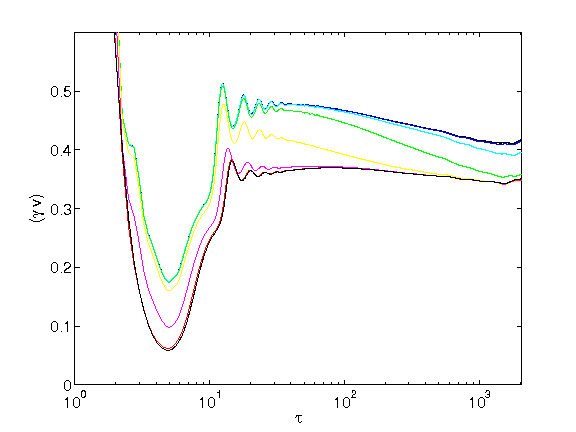}
\end{center}
\caption{\label{RMboxes}The evolution of the dimensionless density ($\rho\tau$, top panel) and $(\gamma v)$ (bottom panel) in $4096^3$ domain wall simulations around the radiation-matter transition. Note that the two black solid lines correspond to the radiation ($\lambda=1/2$) and matter ($\lambda=2/3$) simulations already discussed in Sect.~\protect\ref{Simulation}.}
\end{figure}
%%%%%%%%%%%%%%%%%%%%%%%%%%%%%%%%%%%%%%%%%%%%%%%%%%%%%%%%%%%%%%%%%%%%%%%%%%%%%%%%%%

In this case we ran various sets of simulations with the same parameters and (random) initial conditions that were described in Sect.~\ref{Simulation}, except that the scale factor obeys Eq.~(\ref{ScaleFact}). The requirement of sufficient resolution implies that there is not enough memory available for a single simulation to span the entire transition epoch; instead, various sets of runs were carried out starting at various different conformal times relative to the transition epoch, equally spaced in the logarithm of $\tau_i/\tau_{eq}$.

Figure \ref{RMboxes} (to be compared to Fig. \ref{lambdaboxes}) summarizes the results of these simulations. Note that the two black solid lines correspond to the radiation ($\lambda=1/2$) and matter ($\lambda=2/3$) simulations already discussed in Sect.~\protect\ref{Simulation}. This is an important test of our code: it shows that simulations evolving sufficiently early and sufficiently late in the transition behave exactly like radiation and matter era simulations---as they must. This figure also makes it visually clear that although the 'early' and 'late' simulations reach scaling (since they are effectively evolving with a constant or quasi-constant expansion rate) this is not case for the ones evolving during the transition itself: in that case the effective expansion rate is changing and the network is constantly trying to adapt (as fast as allowed by causality) to these changing conditions. This is clear in the cyan, green and yellow lines in the plots.

\begin{figure}[!]
\begin{center}
\includegraphics[width=3.5in]{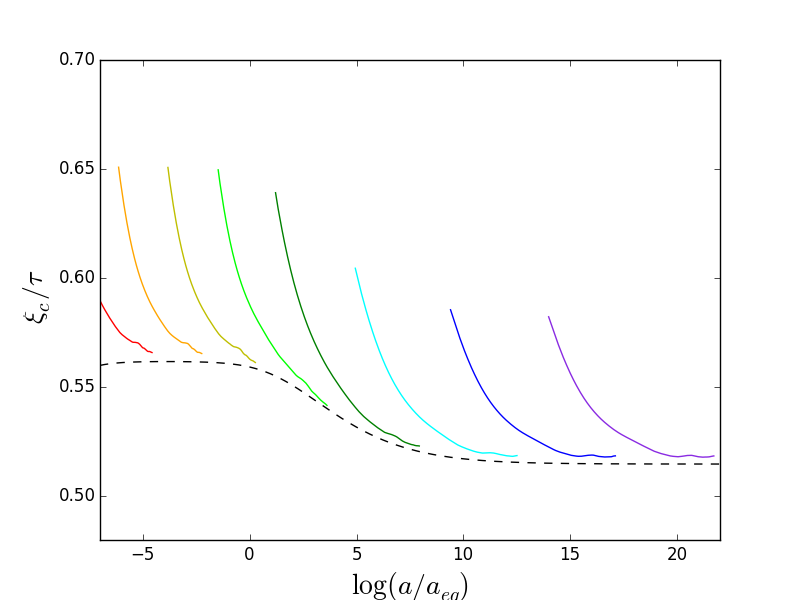}
\includegraphics[width=3.5in]{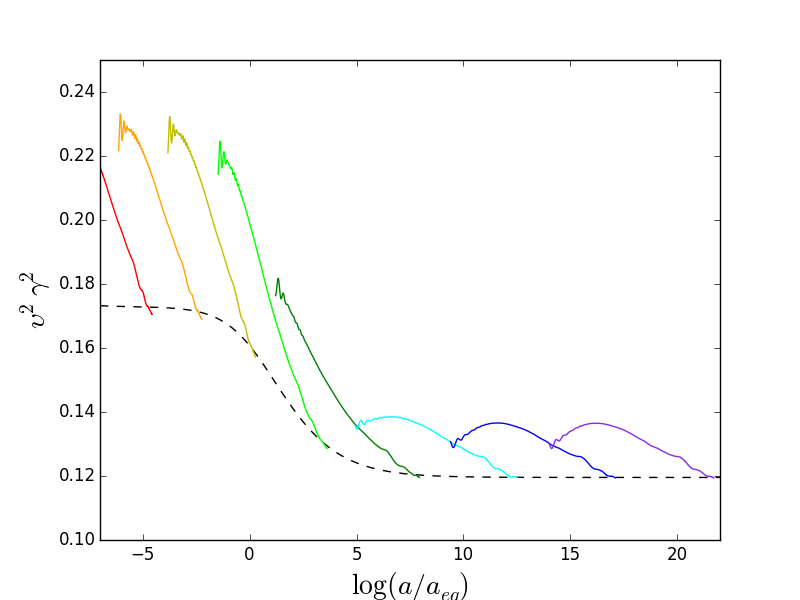}
\end{center}
\caption{\label{Fig:Transition}Evolution of the conformal correlation length divided by conformal time $\xi_c/\tau$ (top panel) and of $(\gamma v)^2$ (bottom panel) during the radiation-matter transition, plotted as a function of the natural logarithm of the scale factor (relative to $a_{eq}$). The simulations are denoted by solid color lines (each line being an average of 10 simulations with random initial conditions) while the prediction of the extended analytic model with the best-fit parameters discussed in the text is shown by the black dashed lines. The plot only includes the dynamic range $20\le\tau\le1500$ of each set of simulations; the earlier part (which is dominated by the initial conditions in the box rather than converging to the attractor solution) and the latter part (which has comparatively poor statistics, as discussed in Sect. III) have been omitted for clarity.} 
\end{figure}

By inserting the scale factor expression (Eq. \ref{ScaleFact}) with the corresponding constants $a_{eq}$ and $\tau_{*}$ in the system of Eqs.~(\ref{AveragedEqSyst3}), we can now compare the dynamics of the extended analytic model and the simulations. This comparison is summarized in Fig.~\ref{Fig:Transition}, where the results of simulations (solid color lines) and the extended analytic model (dashed black line) are compared. It is seen that the analytic model provides an excellent description of the radiation-matter transition.

%%%%%%%%%%%%%%%%%%%%%%%%%%%%%%%%%%%%%%%%%%%%%%%%%%%%%%%%%%%%%%%%%%%%%%%%%%%%%%%%%%
%%%%%%%%%%%%%%%%%%%%%%%%%%%%%%%%%%%%%%%%%%%%%%%%%%%%%%%%%%%%%%%%%%%%%%%%%%%%%%%%%%

%%%%%%%%%%%%%%%%%%%%%%%%%%%%%%%%%%%%%%%%%%%%%%%%%%%%%%%%%%%%%%%%%%%%%%%%%%%%%%%%%%

\section{Conclusions}

In this paper we revisited the evolution of domain wall networks in expanding FLRW universes. We took advantage of recent progress in computing power and hardware to carry out the largest and most extensive set of field theory simulations of domain walls, using the PRS algorithm. We have simulated expanding universes with 12 different fixed expansion rates, as well as sets of simulations which together span the entire radiation-matter transition.

Our simulations allowed us to significantly improve the analytic description of wall network evolution, based on the quantitative VOS model. We have explicitly shown that a constant momentum parameter $k_w$ and chopping parameter $c_w$ cannot fully reproduce the simulations for different expansion rates. As a result of this mismatch, we used phenomenological arguments to introduce an extended model, given by Eqs.~(\ref{AveragedEqSyst3}). In this model the momentum parameter is described by a velocity-dependent function $k(v)$ (Eq. \ref{MomentumParameter}), and there is a generalized energy loss function $F(v)$ (Eq. \ref{ScalarRadiation+Chopping}), which in addition to chopping losses also includes scalar radiation of walls. We did not address the issue of possible losses to gravitational radiation, which is left for subsequent work.

Fitting the phenomenological parameters to the simulations, we found that energy losses due to creation of sphere-like objects are typically subdominant in comparison with scalar radiation, except in the case of fast expansion rates. We have confirmed that the extended analytic model can describe both the fixed expansion rate cases and the transition from the radiation to the matter-dominated era. The latter one is an important test of the model, since the network is not scaling during the transition (while the model parameters were calibrated from fixed expansion rate data in the scaling regime).

In the future it would be interesting to extend this analysis to the case of cosmic strings. In that case the chopping term is known to be more important, but significant and not fully understood differences exist between the results of Goto-Nambu and field theory simulations. As already shown in Ref. \cite{VOS1}, the VOS model allows a direct comparison of the results of both types of simulations, and the recent progress in computing power should permit a clarification of this issue.

%%%%%%%%%%%%%%%%%%%%%%%%%%%%%%%%%%%%%%%%%%%%%%%%%%%%%%%%%%%%%%%%%%%%%%%%%%%%%%%%%%

\begin{acknowledgments}

This work was done in the context of project PTDC/FIS/111725/2009 (FCT, Portugal). CJM is also supported by an FCT Research Professorship, contract reference IF/00064/2012, funded by FCT/MCTES (Portugal) and POPH/FSE (EC). IR is supported by an FCT fellowship (SFRH/BD/52699/2014), under the FCT PD Program PhD::SPACE (PD/00040/2012). IR is grateful for the hospitality of the University of Nottingham, where part of this work was carried out. The work of AA was supported by an Advanced Research Fellowship at the University of Nottingham, United Kingdom.

This work was undertaken on the COSMOS Shared Memory system at DAMTP, University of Cambridge operated on behalf of the STFC DiRAC HPC Facility. This equipment is funded by BIS National E-infrastructure capital grant ST/J005673/1 and STFC grants ST/H008586/1, ST/K00333X/1. The Walls code was extensively modernised by James Briggs (COSMOS IPCC) and John Pennycook (Intel), as described in Ref. \cite{SGI}.

\end{acknowledgments}

\bibliography{walls}
\end{document}